%% file: main.tex
\documentclass[trackchanges,twocolumn]{aastex701}
\usepackage[english]{babel}
\usepackage[utf8]{inputenc}
\usepackage{amsmath}
\usepackage{amsfonts}
\usepackage{amsthm}
\usepackage{amssymb}
\usepackage{graphicx}
\usepackage{xcolor}

\theoremstyle{definition}

\theoremstyle{remark}

\newcommand{\A}{\mathcal{A}}

\def\actaa{\ref@jnl{Acta Astron.}}      % Acta Astronomica
\def\apjl{\textrm{ApJL}}                % Astrophysical Journal, Letters
\def\apjs{\textrm{ApJS}}               % Astrophysical Journal, Supplement
\def\ao{\textrm{Appl.~Opt.}}           % Applied Optics
\def\aap{\textrm{A\&A}}                % Astronomy and Astrophysics
\usepackage[normalem]{ulem}

\begin{document}

\title{ Distinct spin properties and astrophysical origin of low mass binary black holes in gravitational wave data}

\author{Elizabeth Flanagan}
\affiliation{Gravity Exploration Institute, Department of Physics and Astronomy, School of Physical, Chemical and Environmental Sciences, Cardiff University, 5 The Parade, Cardiff, CF24 3AA, United Kingdom}
\email{FlanaganE1@cardiff.ac.uk}

\author{Jakob Stegmann}
\affiliation{Max Planck Institute for Astrophysics, Karl-Schwarzschild-Str. 1, 85748 Garching, Germany}
\email{jstegmann@MPA-Garching.MPG.DE}

\author{Isobel Romero-Shaw}
\affiliation{Gravity Exploration Institute, Department of Physics and Astronomy, School of Physical, Chemical and Environmental Sciences, Cardiff University, 5 The Parade, Cardiff, CF24 3AA, United Kingdom}
\email{Romero-ShawI@cardiff.ac.uk}

\author{Thomas Callister}
\affiliation{Williams College, Williamstown, MA 01267, USA}
\email{tac7@williams.edu}

\author{Aleksandra Olejak}
\affiliation{Max Planck Institute for Astrophysics, Karl-Schwarzschild-Str. 1, 85748 Garching, Germany}
\email{aolejak@MPA-Garching.MPG.DE}

\author{Fabio Antonini}
\affiliation{Gravity Exploration Institute, Department of Physics and Astronomy, School of Physical, Chemical and Environmental Sciences, Cardiff University, 5 The Parade, Cardiff, CF24 3AA, United Kingdom}
\email{AntoniniF@cardiff.ac.uk}

\begin{abstract}
We analyze the effective-spin distribution of binary black hole mergers in GWTC-5.0 as a function of primary black hole mass using hierarchical Bayesian inference. We model the population as a mixture of two spin components separated by a transition mass scale inferred directly from the data. We find strong evidence for a transition at $\tilde{m} = 15.1^{+4.4}_{-3.7}\, M_\odot$. Mock-catalog analyses show that such a transition is unlikely to arise from finite-sample fluctuations of a mass-independent $\chi_{\rm eff}$ population and the posterior predictive distributions of $\chi_{\rm eff}$ inferred below and above the transition are clearly distinct. Below the transition mass, the effective-spin distribution is narrow, peaks at a small positive value $\chi_{\rm eff}>0$, but also shows significant support for negative $\chi_{\rm eff}$. Above the transition, the distribution is broader and its peak shifts to values consistent with $\chi_{\rm eff}\simeq0$, making its support at both positive and negative $\chi_{\rm eff}$ roughly similar. These findings suggest that the dominant merger population concentrated around $10\,M_{\odot}$ is statistically distinct from the rest and that it might arise from a distinct astrophysical process. We show that this low-mass population is broadly consistent with formation from massive stellar multiples in the field: it may arise either from isolated binary star evolution, but only if black hole natal kicks below $\tilde{m}$ are generally very large ($\gtrsim100\,\rm km/s$), or from the dynamical evolution of hierarchical triples. In contrast, isolated binary evolution with standard fallback kick models cannot reproduce the  support for negative $\chi_{\rm eff}$. %The higher-mass black holes likely require additional or different mechanisms.
\end{abstract}

% \maketitle

\section{\label{sec:Intro} Introduction}
\input{intro}

\section{\label{sec:Methods} Methods}
\input{methods}

\section{\label{sec:Results} Results}
\input{results}

\subsection{\label{sec:SimData} Mock Data Analysis}
\input{simulated_data}

\subsection{\label{sec:compare} Comparison to previous work}
\input{comparisonPW}

\section{\label{sec:Discussion} Astrophysical Interpretation}
\input{discussion}

\bibliography{ref}{}
\bibliographystyle{aasjournalv7}

%\clearpage
\section{\label{sec:Appendix} Appendix}
\input{appendix}

%\nocite{*}
% \bibliography{ref}

% \input{sup_mat}

\end{document}

%% file: intro.tex
The origin of merging binary black holes (BBHs) remains one of the central open problems in gravitational-wave astrophysics. Observations from the Advanced LIGO--Virgo--KAGRA network \citep{Abbott:2020niy,2021ApJ...913L...7A, PhysRevX.13.041039,2023PhRvX..13a1048A,LVK_GWTC4_2025,2026arXiv260527225T} have revealed a BBH population with a rich diversity of mass and spin properties, suggesting that multiple astrophysical formation channels may contribute to the observed merger sample \citep{2023PhRvX..13a1048A,2025arXiv250818083T, 2026arXiv260527226T,2022ApJ...928...75H, 2025PhRvL.134a1401A, 2025arXiv250819208M, 2023MNRAS.523.4539K, 2024arXiv240601679P, afroz2025phasespacebinaryblack, afroz2025binaryblackholephase, padhyegurjar2026bbhgenesisdisentanglingbinaryblack, 2026PhRvL.137b1407L, 2026SCPMA..6999562W, 2026arXiv260524281H, alvarez2026effective_spin_gwtc5,Rinaldi:2026nyb,2026arXiv260614472F}. 

The best measured spin combination in gravitational-wave observations is the effective-spin parameter \citep{2008PhRvD..78d4021R, Ajith},
%\begin{equation}
$\chi_{\rm eff} =
({m_1 \chi_1 \cos\theta_1 + m_2 \chi_2 \cos\theta_2})/(m_1+m_2),$
%\end{equation}
 where $m_1$ and $m_2$ are the component black hole masses, $\chi_1$ and $\chi_2$ are the dimensionless spin magnitudes, and $\theta_1$ and $\theta_2$ are the tilt angles between the individual black hole spin vectors and the orbital angular momentum vector.
The effective spin
provides one of the most powerful probes of BBH formation \citep{2018PhRvD..98h3007N, biscoveanu2026decadegravitationalwavemeasurementsblack}, as it encodes information about both black hole spin magnitudes and spin-orbit misalignments while remaining accurately measurable from gravitational-wave signals.
Different formation scenarios predict qualitatively distinct spin properties, ranging from spins preferentially aligned with the orbit for BBH formation from isolated binary star evolution \citep{Gerosa2018,Belczynski2020b}, near-perpendicular spin-orbit orientations from hierarchical triples \citep{Antonini2018,Liu2018,Rodriguez2018,Su2021,2026ApJ..1000L..59S}, to randomly oriented spins in dynamically assembled binaries in star clusters or other dense environments \citep{Rodriguez2016c,2026arXiv260320430A,2026arXiv260621691M}. These spin-orbit orientations lead to different $\chi_{\rm eff}$ distributions that are characterized by a different peak value of $\chi_{\rm eff}$ and ratio between negative and positive values \citep{biscoveanu2026decadegravitationalwavemeasurementsblack}.

Previous analyses of LVK data have reported growing evidence for multiple BBH subpopulations with distinct spin and mass-ratio properties \citep{2023arXiv230401288G, 12025arXiv250602250S, 2022ApJ...941L..39W, 2022ApJ...932L..19B,2021ApJ...922L...5C,2023MNRAS.522..466A, hussain2024hintsspinmagnitudecorrelationsrapidly, ray2024searchingbinaryblackhole, tong2025subpopulationlowmassspinningblack, plunkett2026signaturessubpopulationhierarchicalmergers, banagiri2025evidencesubpopulationsmergingbinary,2026arXiv260614472F} and identified a transition to a broad $\chi_{\rm eff}$ distribution for mergers with primary masses $m_1 \gtrsim 45~M_\odot$ \citep{ 2025PhRvL.134a1401A, nxnr-pdyx, antonini2026a}. This high-mass population has been interpreted as being dominated by hierarchical, second-generation mergers, in which rapidly spinning black holes are remnants of previous mergers assembled dynamically in dense stellar clusters \citep{2006ApJ...637..937O,2016ApJ...831..187A,2019PhRvD.100d3027R}. \cite{2025arXiv251105316T} further presented evidence for an additional low-mass subpopulation with large, isotropically distributed spins, including the exceptional events GW241011\_233834 and GW241110\_124123 (henceforth GW241011 and GW241110 respectively) \citep{2025ApJ...993L..21A}, which is also consistent with a hierarchical merger origin.
Evidence for multiple distinct BBH populations has also emerged from analyses of the joint mass-ratio, spin and redshift distributions \citep[e.g.,][]{banagiri2025evidencesubpopulationsmergingbinary,2026arXiv260317987R}. 
%Ref.~\cite{2026arXiv260527226T} analyzed the GWTC-5 catalog \cite{2026arXiv260527223T, 2026arXiv260527224T, 2026arXiv260527225T} with a three-transition model for the mass dependence of the effective-spin distribution.  The inferred transition masses have partially overlapping marginal posteriors, making it unclear whether the new data require three distinct physical transitions or instead favor a more general change in spin properties over a broad mass range. 
Furthermore, previous works  identified weak evidence for the $\chi_{\rm eff}$ distribution of the $10$M$_{\odot}$ peak being distinct from the broader BBH population using earlier GW catalogs \citep{2023arXiv230401288G, 2026arXiv260317987R, 2026arXiv260527226T, 2026arXiv260401420L}. In particular, \cite{2026arXiv260317987R} reported evidence for the $10$M$_{\odot}$ peak having a $\chi_{\rm eff}$ distribution with a positive mean, while the remaining distributions  were consistent with being centered near zero.
These previous works motivate the  questions addressed here: is there a single dominant low-mass scale at which the
$\chi_{\rm eff}$ distribution changes, and if so, what is the origin of this transition?

% We follow the approach  developed in Ref.\cite{2025PhRvL.134a1401A} to investigate the mass dependence of the BBH spin distribution using flexible non-parametric population models applied to GWTC-5. We model the population as a mixture of two independent effective-spin distributions separated by a transition mass scale inferred directly from the data. We find a  transition at $\tilde{m}\simeq 15~M_{\odot}$.  
 %These results provide evidence that the dominant low-mass BBH population is statistically distinct from the rest of the merger population.  A companion analysis supports this picture by showing that the joint mass-ratio and effective-spin distribution across the full mass range is inconsistent with that of the low-mass population alone~\citep{2026arXiv260614472F}.
 %We discuss the  properties of this distribution in relation to astrophysical formation scenarios, and show under which assumptions isolated binary and triple evolution provide a possible explanation for this population.

\bigskip

%% file: methods.tex
% \noindent{\emph {Methods}}~
Using the latest gravitational-wave catalog, GWTC-5.0 \citep{2026arXiv260527225T}, we perform a hierarchical Bayesian analysis of the binary black hole population. Following previous analyses~\cite[e.g.,][]{2025arXiv250818083T}, we restrict our sample to events with false alarm rate $\mathrm{FAR}<1~{\rm yr}^{-1}$, yielding a total of 259 mergers. Selection biases are incorporated using recovered injections from dedicated injection campaigns \citep{PhysRevX.13.041039, 44x3-hv3y}.

% We analyze the BBH population with flexible Gaussian-process population models. The primary-mass spectrum is modeled non-parametrically with a Gaussian process, while the conditional secondary-mass distribution is described by a power-law index that varies with primary mass also through a Gaussian-process model. The redshift evolution is parameterized as a power law in $(1+z)$.

We analyze the BBH population with flexible non-parametric models. In particular, the primary mass distribution and the power-law index of the conditional secondary-mass distribution, are modeled as functions of primary mass using Gaussian processes. A more detailed description of these models is given in the Supplementary Materials (SM).

In our main analysis, we model the $\chi_{\rm eff}$ distribution as a mixture of two spin populations separated by a transition mass scale $\tilde{m}$, which is inferred directly from the data. The effective-spin distribution conditioned on primary mass is written as
\begin{eqnarray}
p(\chi_{\rm eff}\mid m_1)
=
p_{\rm low}(\chi_{\rm eff}\mid m_1)\,[1-\zeta(m_1)]
+\nonumber \\
p_{\rm high}(\chi_{\rm eff}\mid m_1)\,\zeta(m_1),
\end{eqnarray}
where $\zeta(m_1)$ is a sigmoid mixing function that transitions exponentially around $\tilde{m}$, 
\begin{equation}
    \zeta(m_1)
=
\frac{1}{1+e^{{-(m_1-\tilde{m}_{})/M_\odot}}}.
\end{equation}
Each spin component is modeled non-parametrically,  allowing the data to determine both the shape and support of the $\chi_{\rm eff}$ distribution without imposing a fixed functional form:
\begin{equation}
p_{\rm i}(\chi_{\rm eff}\mid m_1)
=
\frac{
\mathcal{H}_i(\chi_{\rm eff})
\,e^{\Theta_i(\chi_{\rm eff})}
}{
\int_{-1}^{1}
\mathcal{H}_i(\chi_{\rm eff})
\,e^{\Theta_i(\chi_{\rm eff})}
\,d\chi_{\rm eff}
},
\end{equation}
where \(i\in\{\mathrm{low},\mathrm{high}\}\) labels the spin distribution either above
or below $\tilde{m}$. For each
component, \(\Theta_i(\chi_{\rm eff})\) is modeled as a Gaussian process, while
\(H_i(\chi_{\rm eff})\) is a Heaviside window function that defines the
allowed support of the distribution:
$\mathcal{H}_i(\chi_{\rm eff})
=1$ for $
 \chi_{{\rm min},i}\leq \chi_{\rm eff}\leq \chi_{{\rm max},i}$ or 
$\mathcal{H}_i(\chi_{\rm eff})
= 0$ otherwise. This construction enables the identification of localized features and statistically distinct spin populations associated with different regions of the BBH mass spectrum.

To understand how our results depend on the adopted population model, we consider an additional stronger parametric model where below the mass transition the spin distribution is a truncated normal $\mathcal{N}(\chi_{\rm eff};\mu_i,\sigma_i)$ and  above the transition is a mixture of a truncated
normal and a uniform,
%\begin{equation}
$
\theta\,\mathcal{N}(\chi_{\rm eff};\mu_i,\sigma_i)
+
(1-\theta)\,\mathcal{U}(\chi_{\rm eff};\chi_{{\rm min}},\chi_{{\rm max}}).$
%\end{equation}
Here
$\mathcal{N}$ is truncated to $-1 \leq\chi_{\rm eff}\leq 1$ and  $\chi_{{\rm min}}$ and $\chi_{{\rm max}}$ are the bound of the uniform distributions that are independent from each other and are also inferred from the data.

The hyperparameter priors entering our models are listed in the SM and are chosen to be weakly informative. The only exception is the prior on the transition mass, $\tilde m$, which we take to be uniform over the interval $3$--$35~M_\odot$. This range deliberately excludes the $\sim45~M_\odot$ transition identified in previous work \citep{2025PhRvL.134a1401A}, enabling us to test specifically for an additional transition at lower primary masses.

A similar model was employed in our companion analysis \citep{2026arXiv260614472F}, where instead of letting the mass transitions be a free parameter, multiple transitions were manually selected based on the mass ratio distribution of the population. 
Here, our model is designed to identify the leading-order low-mass transition in the mass dependence of the \(\chi_{\rm eff}\) distribution. This does not however exclude additional structure above \(\sim15\,M_\odot\).

\bigskip

%% file: results.tex
\begin{figure}
\centering
\resizebox{0.49\textwidth}{!}{\includegraphics{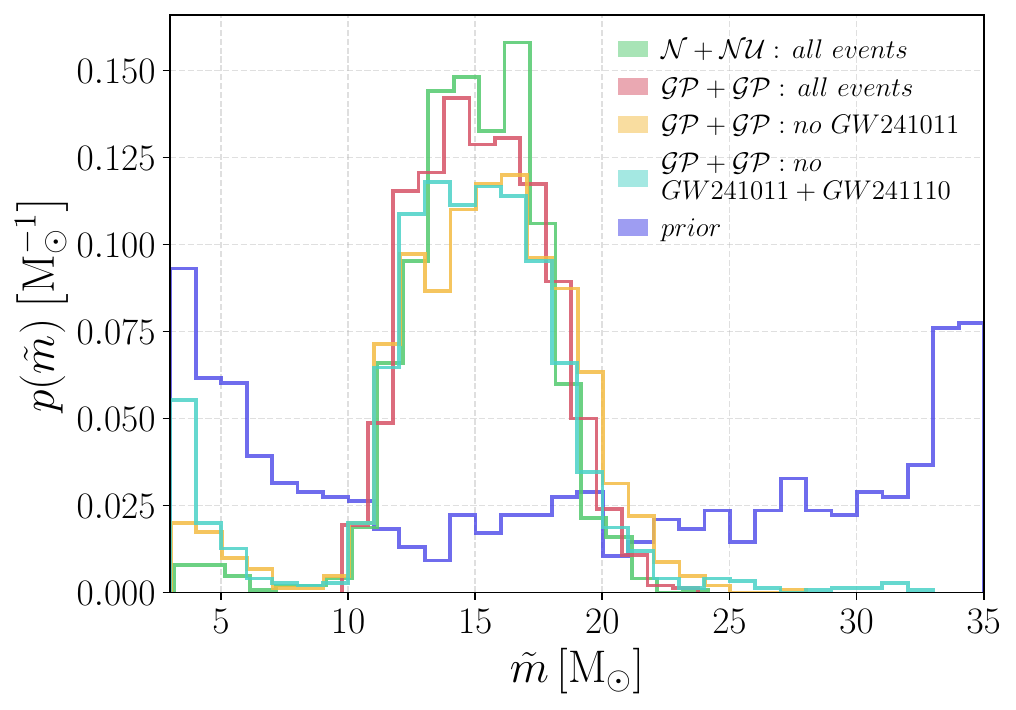}}
\caption{Mass transition posteriors inferred using the $\mathcal{N} + \mathcal{NU}$ model and the $\mathcal{GP} + \mathcal{GP}$. Also shown is the mass transition inferred when using the $\mathcal{GP} + \mathcal{GP}$ model and removing both GW241011 and GW241110. The prior is constructed with draws from the mass transition prior which pass the $N_{\rm eff}$ cutoffs.}
\label{fig:mCuts_posterior}
\end{figure}

% \noindent{\emph {Results}}
The posterior distributions for the recovered mass transitions $\tilde{m}$ are shown in Fig.~\ref{fig:mCuts_posterior}. For our non-parametric ($\mathcal{GP} + \mathcal{GP}$) and parametric ($\mathcal{N} + \mathcal{NU}$) models, we infer $\tilde{m} = 15.1^{+4.4}_{-3.7}\,M_\odot$ and $ 15.1^{+3.8}_{-3.9}\,M_\odot$ respectively (hereafter, quoted uncertainties denote the $90\%$ credible interval).
The posterior predictive distributions indicate that this behavior is robust across posterior realizations. 

We use a leave-one-out analysis to show that the result is not driven by a small subset of influential events. We focus in particular on GW241110 and GW241011 as they are low-mass events with substantial support at positive or negative $\chi_{\rm eff}$ \citep{2025ApJ...993L..21A}, and therefore are natural candidates for driving the inferred broadening of the high-mass spin distribution. Removing GW241110, or both  GW241011 and GW241110 events from the sample leaves the inferred transitional mass essentially unchanged (see Fig.~\ref{fig:mCuts_posterior}).

% The posterior distribution for $\tilde m$ recovered by our model is shown in Fig.~\ref{fig:mCuts_posterior}.
% We infer
% $
% \tilde m = 15^{+4}_{-4} M_\odot ,
% $
% (where, hereafter, quoted uncertainties denote the $90\%$ credible interval).%indicating strong evidence for a transition in the spin properties of merging BBHs at low masses.  

\begin{figure}
\centering
\resizebox{0.5\textwidth}{!}{\includegraphics{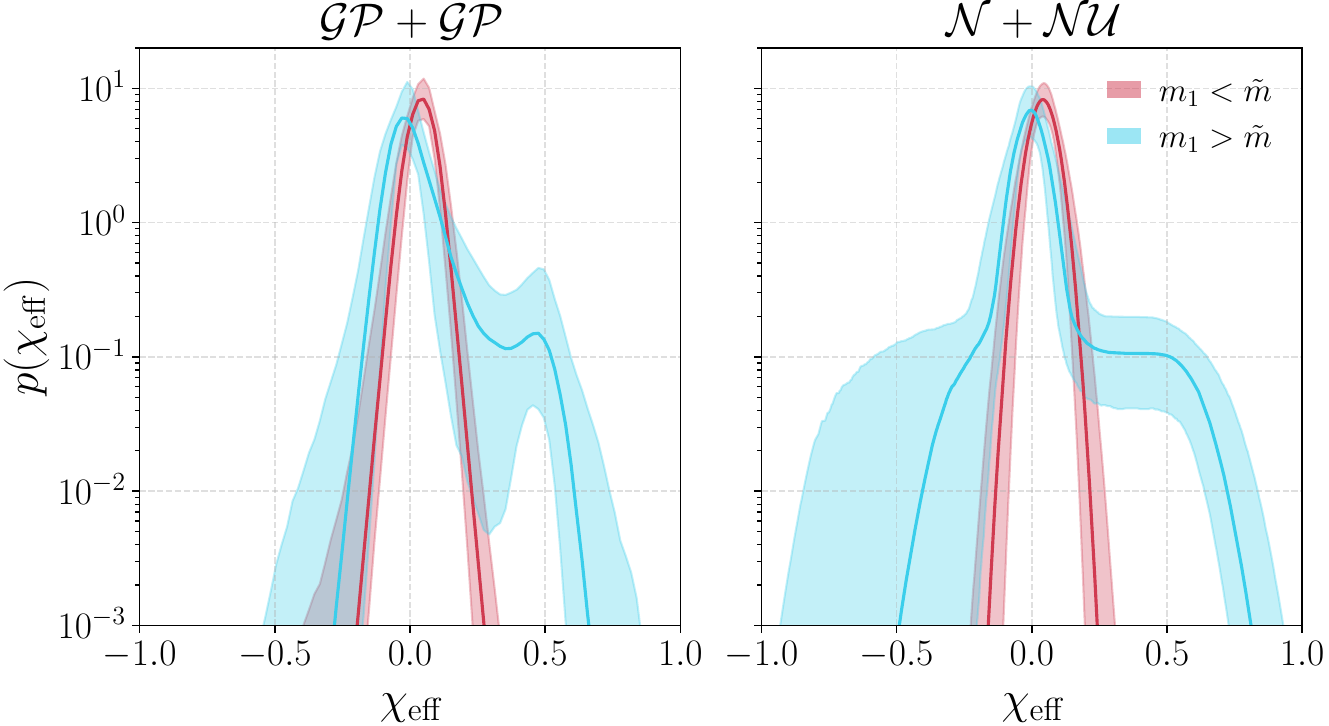}}
\caption{The effective spin posteriors for both the non-parametric $\mathcal{GP} + \mathcal{GP}$ (left) and parametric $\mathcal{N} + \mathcal{NU}$ (right) models. For the effective spin posteriors, the shaded area represents the $90 \%$ confidence interval and the solid lines are the recovered median of the distribution.
}
\label{fig:posterior}
\end{figure}

\begin{table*}
\centering
\begin{tabular*}{\textwidth}{@{\extracolsep{\fill}} l c c c c c }
\hline
Model & Mean & Median & Mode & Skew & Width\\ 
\hline
$\mathcal{GP}+\mathcal{GP}: \, m_1 < \tilde{m}$ & $0.041_{-0.017}^{+0.018}$ & $0.032_{-0.016}^{+0.017}$ & $0.051_{-0.020}^{+0.020}$ & $-0.034_{-0.33}^{+0.28}$ & $0.047_{-0.010}^{+0.019}$\\
$\mathcal{GP}+\mathcal{GP}: \, m_1 > \tilde{m}$ & $0.015_{-0.022}^{+0.022}$ & $-0.019_{-0.024}^{+0.022}$ & $-0.030_{-0.040}^{+0.040}$ & $0.36_{-0.32}^{+0.25}$ & $0.12_{-0.022}^{+0.026}$ \\
$\mathcal{N}+\mathcal{NU}: \, m_1 < \tilde{m}$ & $0.041_{-0.015}^{+0.018}$ & $0.031_{-0.015}^{+0.018}$ & $0.051_{-0.020}^{+0.00}$ & $-0.010_{-0.19}^{+0.21}$ & $0.047_{-0.012}^{+0.016}$\\
$\mathcal{N}+\mathcal{NU}: \, m_1 > \tilde{m}$ & $0.010_{-0.024}^{+0.022}$ & $-0.013_{-0.023}^{+0.024}$ & $-0.010_{-0.020}^{+0.020}$ & $0.13_{-0.17}^{+0.12}$ & $0.12_{-0.041}^{+0.059}$\\
% $\mathcal{GP}+\mathcal{GP}: \, m_1 < \tilde{m}$ & $0.04^{+0.02}_{-0.02}$ & $0.03^{+0.02}_{-0.02}$ & $0.05^{+0.02}_{-0.02}$ & $-0.03^{+0.28}_{-0.33}$ & $0.05^{+0.01}_{-0.02}$\\
% $\mathcal{GP}+\mathcal{GP}: \, m_1 > \tilde{m}$ & $0.02^{+0.02}_{-0.02}$ & $-0.02^{+0.02}_{-0.02}$ & $-0.03^{+0.04}_{-0.04}$ & $0.36^{+0.25}_{-0.32}$ & $0.12^{+0.02}_{-0.03}$\\
% $\mathcal{N}+\mathcal{NU}: \, m_1 < \tilde{m}$ & $0.04^{+0.02}_{-0.02}$ & $0.03^{+0.02}_{-0.02}$ & $0.05^{+0.00}_{-0.02}$ & $-0.01^{+0.21}_{-0.19}$ & $0.05^{-0.01}_{+0.02}$\\
% $\mathcal{N}+\mathcal{NU}: \, m_1 > \tilde{m}$ & $0.01^{+0.02}_{-0.02}$ & $-0.01^{+0.02}_{-0.02}$ & $-0.01^{+0.02}_{-0.02}$ & $0.13^{+0.12}_{-0.17}$ & $0.12^{-0.04}_{+0.06}$\\
\hline
\end{tabular*}
\caption{Summary statistics of the effective spin distribution above and below the inferred mass transition for both models.}
\label{tab:summarystats}
\end{table*}

If the data favored a broad or prior-limited posterior on $\tilde m$, this would suggest that no preferred transition scale is required to describe the observed BBH population, and/or that the $\chi_{\rm eff}$ distribution varies smoothly across the full mass range. Instead, we find that $\tilde m$ is tightly localized by the data, providing strong evidence for a change in the effective-spin distribution at a characteristic primary mass. 
This indicates that BBHs below and above $\tilde m$ are likely to be statistically described by different spin populations.

In Fig.~\ref{fig:posterior} we show the inferred $\chi_{\rm eff}$ distributions below and above the transition mass for the two adopted population models.
Below $\tilde m$, the effective-spin distribution is narrow and peaks at positive values of $\chi_{\rm eff}$.
The distribution exhibits significant support for negative effective spins but is concentrated around a positive value $\chi_{\rm eff}\simeq 0.03$. The inferred width of the distribution is $\sigma_{\chi}\simeq 0.05$, indicating a relatively homogeneous low-spin population.
The high-mass component exhibits a significantly larger variance, a peak that is  consistent with zero, and more equal support for negative and positive values of $\chi_{\rm eff}$.

Summary statistics for the effective spin distribution are reported in Table \ref{tab:summarystats} for both population models. In comparing the distributions above and below the mass transition, we see a statistically significant difference in the median, mode, and width. For each model, both the mode and the width of each subpopulation disagree at a 99\% confidence level, while the medians disagree at a $>99\%$ confidence level, further indicating that these two regions are statistically distinct from one another. We also compute the
skew of the distribution with respect to its mode and 
see weak evidence for positive skewness in the distribution above the mass transition in both models. Additionally, below the transition, both models find the skewness being consistent with zero. This suggests that the positive skewness identified in previous analyses 
\citep{2026arXiv260527226T, Banagiri_2025}, if present, is most likely a property of the population above $\simeq 15M_\odot$ in agreement with concurrent work \citep{alvarez2026effective_spin_gwtc5,Rinaldi:2026nyb,2026arXiv260614472F}.

\bigskip

%% file: simulated_data.tex
% \noindent{\emph {Mock data analysis}}~
We validate that our result could not be spuriously achieved through an analysis of a population with a skewnormal $\chi_\mathrm{eff}$ distribution consistent with that inferred in GWTC-5.0 \citep{2026arXiv260527226T}. Such test is  necessary in order to draw robust conclusions from the gravitational-wave catalog \citep{2026PhRvD.113j3021M, GWMockCat}.
Using mock parameter estimation simulation code \texttt{GWMockCat} \citep{GWMockCat}, updated to include simulated $\chi_\mathrm{eff}$ samples \citep{IRS_inprep}, we simulate $100$ realizations of posterior distributions for $259$ events, each with $1000$ posterior samples. We also produce $5\times10^7$ found injections, assuming O4 detector sensitivity, to incorporate  observational selection effects.\footnote{\textbf{We use the noise curve stored in \texttt{aligo\_O4high.txt} at \href{https://dcc.ligo.org/LIGO-T2000012/public}{https://dcc.ligo.org/LIGO-T2000012/public}.}} Our found injections are shown in Figure \ref{fig:simulated_data} in the SM, along with further details about our assumed parameter distributions and an example mock posterior compared with a real observation. Our base population has no correlation between $\chi_\mathrm{eff}$ and $m_1$, i.e., there is no $\tilde{m}$ at which the $\chi_\mathrm{eff}$ distribution changes.

We analyze these $100$ population realizations with the  $\mathcal{N}+\mathcal{NU}$ model described in Table \ref{tab:summarystats}. 
In all cases we find that the predictive posterior distribution of $\tilde{m}$ has some finite support at the boundaries of the prior (i.e., at $3~M_\odot$, $35~M_\odot$, or both), implying that the absence of a transition in the mock data cannot be ruled out, consistent with its absence in the  injected population. 
This is different from the analysis of the real GW data where the  posterior distribution of $\tilde{m}$ shows no support at the prior boundaries.

In all but two cases  the posterior on $\tilde{m}$ rails at one or both edges of the prior boundary with small or no support in between.
 This behavior indicates that the data do not identify a preferred transition scale in these mock realizations. %Instead, the transition parameter is effectively unconstrained, and the model places the division at the prior boundaries rather than recovering a robust mass-dependent spin feature.
In the two remaining cases 
we infer $\tilde{m}=15.54^{+10}_{-5}$~M$_\odot$, and
$\tilde{m}=13.54^{+3}_{-5}$~M$_\odot$.
These mock realizations show that a mass-independent spin population can occasionally produce a transition-like feature. However, the feature is not entirely comparable to the real-data result. The inferred transition has a broader uncertainty than in the real data, and the two effective spin distributions are more weakly separated, with  differences being statistically significant only outside the bulk of the population $|\chi_{\rm eff}|\gtrsim 0.3$ (see SM). We investigate one of these population realizations further in the SM, where we plot the mock population draws above and below $15$~M$_\odot$, plus the posteriors of the simulated events with the most and least extreme $\chi_\mathrm{eff}$. As can be seen in Figs. \ref{fig:simulated_data} and \ref{fig:examples}, this population has both outliers at large $\chi_\mathrm{eff}$ due to Poisson noise, and posteriors peaking at more extreme $\chi_\mathrm{eff}$ than injected due to Gaussian noise. We believe that the inferred transition occurs in these cases due to unlucky Poisson draws from the population, plus unlucky shifts in the posteriors due to Gaussian noise.

We conclude that our results are unlikely to be explained by being a particular realization of a random draw from a skewnormal $\chi_\mathrm{eff}$ distribution similar to that inferred in GWTC-5.0.
\bigskip

%% file: comparisonPW.tex
Our results should be interpreted in the context of growing evidence for multiple BBH subpopulations. \citet{banagiri2025evidencesubpopulationsmergingbinary} found support for three BBH subpopulations separated by  transitions in primary mass: systems with $m_1\lesssim 28~M_{\odot}$ have nearly flat mass-ratio distributions and small spins, binaries with $28\lesssim m_1\lesssim 40~M_{\odot}$ strongly favor equal masses, and systems with $m_1\gtrsim 40~M_{\odot}$ show support for larger spins and possibly asymmetric mass ratios. \citet{2026arXiv260317987R} presented a different interpretation in which the observed BBH population is composed of three overlapping components associated with different formation channels, with the $\sim 10~M_{\odot}$ and $\sim 35~M_{\odot}$ mass-spectrum features exhibiting distinct spin-alignment, mass-ratio, and redshift-evolution properties.

Some recent studies have  reported a broadening  of the $\chi_{\rm eff}$ distribution in the  mass interval $m_1\simeq 13$--$20~M_{\odot}$ \citep{2025arXiv251105316T,plunkett2026signaturessubpopulationhierarchicalmergers}. This feature was not statistically required in GWTC-3 and GWTC-4.0~\citep{nxnr-pdyx,antonini2026a}, but becomes more significant with the enlarged GWTC-5.0 catalog \citep{2026arXiv260527226T,2026arXiv260527223T,2026arXiv260527224T,2026arXiv260527225T}.
Similarly,  \cite{2026arXiv260527226T} show the possibility of two transitions existing in this low-mass region.  
Whilst our analysis cannot demonstrate the effective spin distribution temporarily departing from, and returning to, the behavior of the broader population, we do find that the dominant low-mass population below $m_1\simeq 15~M_{\odot}$ has a $\chi_{\rm eff}$ distribution that is statistically distinct from the rest of the catalog.
This interpretation is supported by our companion analysis  \citep{2026arXiv260614472F} which finds that $m_1\lesssim 15~M_{\odot}$ is the only mass range where the peak of $p(\chi_{\rm eff})$ is inconsistent with zero, and that the same region also differs in mass ratio and in the width of the effective-spin distribution. 
Additionally, our interpretation supports previous works that also identify the $\chi_{\rm eff}$ distribution in the low-mass regime as potentially distinct from the broader population \cite{banagiri2025evidencesubpopulationsmergingbinary, 2026arXiv260317987R}.
Our result does not imply that there is only one transition in this mass range. If multiple mass scales exist in this region around $\sim 13-20$ M$_{\odot}$, as suggested by previous work, then our result is averaging over these transitions. Nonetheless, the lower-mass population below our inferred $\tilde{m}$ is still statistically distinct from the rest, as is seen in the effective spin.

% Our result is qualitatively different from identifying a narrow mass interval in which the spin distribution temporarily departs from, and then returns to, the behavior of the broader population. Instead, we find that the dominant low-mass population below $m_1\simeq 15~M_{\odot}$ has a $\chi_{\rm eff}$ distribution that is statistically distinct from the rest of the catalog. 

\iffalse
\cite{alvarez2026effective_spin_gwtc5} recently modeled $p(\chi_{\rm eff},m_1)$ as a mixture of a Gaussian-like bulk
with mass-independent spin parameters and an additional broad,
subdominant component whose contribution and shape vary with
primary mass. {\bf Our analysis complements this
decomposition.} We find that the main peak of
$p(\chi_{\rm eff}\mid m_1)$ itself changes with $m_1$, so the
mass dependence might not be fully captured by adding a
mass-dependent residual component on top of a common bulk.
Thus, while both analyses find evidence for mass-dependent spin
structure, they differ in whether that dependence is attributed
primarily to the dominant spin population or to an additional
residual component.
\fi

 Taken together, the enlarged GWTC-5.0 catalog, the agreement with other analyses  \citep[e.g.,][]{2026arXiv260317987R,2026arXiv260614472F}, and the mock-catalog validation presented here provide  support to the hypothesis that the dominant low-mass BBH population might be a distinct component of the detected population, motivating the astrophysical interpretation developed below.

%% file: discussion.tex
% \noindent{\emph {Astrophysical modeling}}
%Motivate 10 Msun peak from stellar collapse
The distinct spin properties of the low-mass subpopulation ($m_1\lesssim\tilde{m}\simeq15\,M_\odot$) allows us to scrutinize previous formation scenarios that have been proposed to explain its astrophysical origin. The low-mass population is particularly important since it encloses the global peak of the mass distribution around $\sim10\,M_\odot$ which dominates the total population of BBH mergers. While BH dynamics in dense environments (e.g., globular and nuclear clusters) is thought to yield relatively massive mergers \citep{Askar2017,2019PhRvD.100d3027R,2021MNRAS.505..339M,2023MNRAS.522..466A,2023MNRAS.520.5259A,2026arXiv260320430A} and formation in young--massive star cluster might even explain the $10M_\odot$ peak \citep{2026arXiv260320430A,2026ApJ...997..267Y}, the most common interpretation for the origin of the low-mass BH population is based on the large observed fraction of isolated massive stellar binaries and higher-order multiples in the field \citep{Moe2017}. In these systems, binary stripping 
%- particularly envelope stripping - combined with 
and core-collapse physics 
%that favors either successful supernova explosions producing neutron stars or direct collapse to BHs, 
may result in a compact-object mass distribution peaking around $\sim 10\,M_\odot$ \citep[e.g.,][]{Fryer12, Belczynski2012a, Sukhold2016, Schneider2021, Olejak2022, vanSon2022,BanerjeOlejak2024, 2025A&A...694A.186G, galaudage2026compactnesspeakssubpopulationsprobing}.

%Yet the  mechanism that produces the observed BBH mergers remains unclear since massive stellar multiples offer two fundamentally different pathways to form close BH pairs. On the one hand, close isolated binary stars could interact with one another through stable or unstable mass transfer and leave behind a BBH which is close enough to merge due to gravitational-wave emission \cite{Belczynski2020b, Gallegos-Garcia2021,Bavera2020, Olejak2021,Klencki2021,vanSon2022,Xu2025,Broekgaarden2026, Sgalletta2026}. On the other hand, they may result from the evolution of hierarchical stellar triples or higher-order configurations \cite{Silsbee2017,Antonini2017,Grishin2018,Liu2018,Antonini2018,Rodriguez2018,Mangipudi2022,Stegmann2022,Stegmann2022b,2025A&A...699A.272V,Dorozsmai2025,Stegmann2025,2026ApJ..1000L..59S}. In triples,  the merger of a BBH formed in the inner binary is driven by the gravitational perturbation from a distant tertiary companion that drives large-amplitude eccentricity oscillations through the Lidov-Kozai effect \cite{Zeipel1910,Lidov1962,Kozai1962,Naoz2016}.

\begin{figure}
    \centering
    \includegraphics[width=\linewidth]{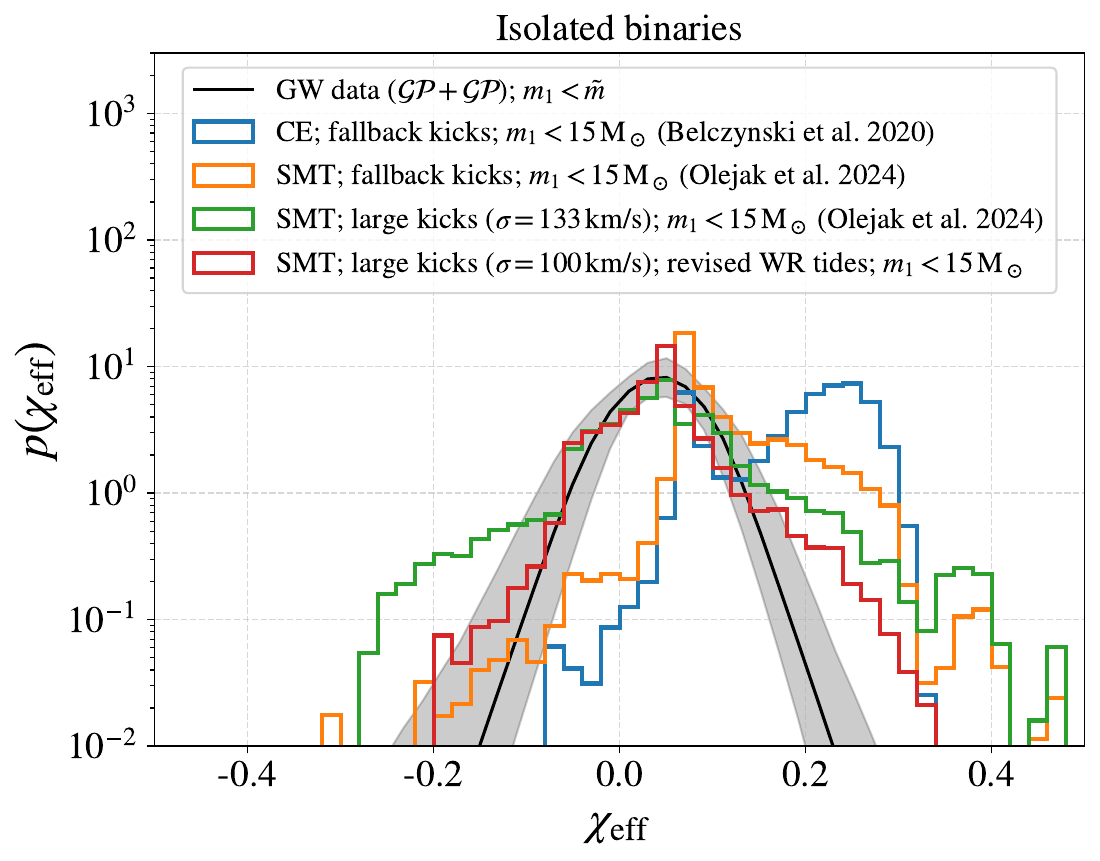}
    \caption{Comparison between the $\chi_{\rm eff}$ distribution inferred for the low-mass subpopulation ($m_1 < \tilde{m}$) using the non-parametric $\mathcal{GP}+\mathcal{GP}$ model and predictions from isolated binary evolution models. The blue curve shows model M30.B from \cite{Belczynski2020b}, in which BBH mergers form through common-envelope evolution, and BHs receive fallback-modulated natal kicks. The orange curve corresponds to the standard stable mass-transfer model from \cite{Olejak2024}, which also adopts fallback-modulated natal kicks. The green curve shows a variant of the stable mass-transfer model from ~\cite{Olejak2024} but with high natal kicks drawn from a Maxwellian distribution with dispersion $\sigma = 133\,\mathrm{km\,s^{-1}}$ and no fallback reduction. The red curve shows a new model with high natal kicks drawn from a Maxwellian distribution with dispersion $\sigma = 100\,\mathrm{km\,s^{-1}}$, also without fallback reduction, with reduced tidal spin-up in BH - helium core binaries. For each model, only BBH mergers with primary masses $m_1 < 15\,M_\odot$ are included.}
    \label{fig:astro_model_comparison_binaries_Aleksandra}
\end{figure}

%\begin{figure}
%    \centering
%    \includegraphics[width=\linewidth]{astro_model_comparison_triples.pdf}
%    \caption{Comparison between the $\chi_{\rm eff}$ distribution inferred for the low-mass subpopulation ($m_1<\tilde{m}$) in the non-parametric $\mathcal{GP} + \mathcal{GP}$ analysis and the expected distribution from BBH mergers that form in hierarchical triples. For the latter, we show the simulation outcomes of a triple population synthesis study that assumes the standard fallback model for natal kicks at BH formation. The BH component spin magnitudes  are independently drawn from a truncated normal distribution between $\chi_{\rm min}=0$ and $\chi_{\rm max}=1$ with \textbf{$\mu=0.1$ and $\sigma=0.25$}. We only show the subset of mergers with $m_1<15\,\rm M_\odot$}
%    \label{fig:astro_model_comparison_triples}
%\end{figure}

\begin{figure}
    \centering
    \includegraphics[width=\linewidth]{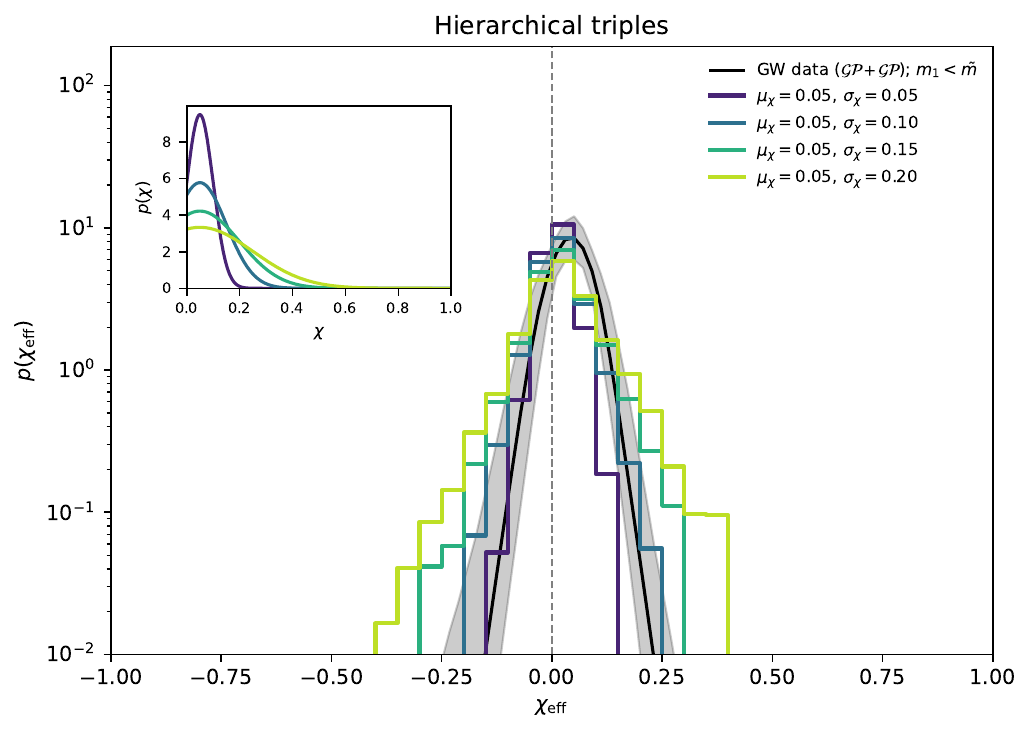}
    \caption{Comparison between the $\chi_{\rm eff}$ distribution inferred for the low-mass subpopulation ($m_1<\tilde{m}$) in the non-parametric $\mathcal{GP} + \mathcal{GP}$ analysis and the expected distribution from BBH mergers that form in hierarchical triples. For the latter, we show the simulation outcomes of a triple population synthesis study that assumes the standard fallback model for natal kicks at BH formation. The BH component spin magnitudes  are independently drawn from a truncated normal distribution between $\chi_{\rm min}=0$ and $\chi_{\rm max}=1$ with $\mu=0.05$ and $\sigma$ ranging from 0.05 to 0.20 (coloured lines). The inset shows the corresponding spin magnitude distributions. We only show the subset of mergers with $m_1<15\,\rm M_\odot$}
    \label{fig:astro_model_comparison_triples}
\end{figure}

While binary stripping combined with direct core-collapse BH formation can reproduce the $\sim 10\,M_{\odot}$ peak of the mass distribution, standard isolated binary evolution models face difficulties in reproducing the negative $\chi_{\rm eff}$ fraction inferred from GW observations. In Fig.~\ref{fig:astro_model_comparison_binaries_Aleksandra}, we show the predicted $\chi_{\rm eff}$ distribution from several representative examples of isolated binary evolution models. The two models from \cite{Belczynski2020b} (blue) and \cite{Olejak2024} (orange) incorporate standard assumptions on BH formation in which BHs receive relatively small natal kicks at birth due to fallback suppression of the explosion asymmetry \citep{Fryer12}. Since BHs are assumed to be born with low natal spins \citep{Fuller2019} \citep[except cases of tidally spun-up second-born BH progenitor;][]{OlejakBelczynski2021}, these models preferentially produce small positive values of $\chi_{\rm eff}$. While such models are commonly invoked to explain the formation of BBH mergers---especially near the $10\,M_{\odot}$ peak  \citep[e.g.,][]{Zevin2021,2025arXiv250818083T,2026arXiv260317987R,QiuCheng2026,alvarez2026effective_spin_gwtc5}---comparison with the inferred distribution from LVK data (black) shows that standard isolated binary evolution models fail to reproduce the significant fraction of negative $\chi_{\rm eff}$ values inferred from GW observations.

In classical isolated binary evolution, due to expected preference for initial alignment, subsequent binary interactions, and low natal kicks, BBHs are expected to form with spins closely aligned with the orbital angular momentum, corresponding to spin–orbit angles $\cos\theta_{1(2)} \simeq 1$ \citep{Farr2017, Belczynski2020b, Bavera2020, OlejakBelczynski2021}. A significant fraction of negative values of $\cos\theta_{1(2)}$, and hence negative $\chi_{\rm eff}$, is  expected if BHs receive sufficiently larger natal kicks than in standard fallback models, that could significantly tilt the orbital angular momentum with respect to the component spins \citep{Gerosa2018}.  Models in which BHs receive larger natal kicks %that lead to a significant population of binaries with negative $\chi_{\rm eff}$ 
are included in Figure~\ref{fig:astro_model_comparison_binaries_Aleksandra} (red and green). These models assume BH natal kicks drawn from Maxwellian distributions with dispersions $\sigma \sim 100\,\mathrm{km\,s^{-1}}$ and without fallback suppression.\footnote{The model with $\sigma = 100\,\mathrm{km\,s^{-1}}$ (red line, Fig. \ref{fig:astro_model_comparison_binaries_Aleksandra}) additionally differs from the $\sigma = 133\,\mathrm{km\,s^{-1}}$ model (green line, Fig. \ref{fig:astro_model_comparison_binaries_Aleksandra}) by reduced tidal spin-up in stripped helium core--BH binaries, accounting for orbital widening due to stellar winds during the phase.} These models provide a better match to both the observed fraction of negative $\chi_{\rm eff}$ systems and the inferred location of the peak in the $\chi_{\rm eff}$ distribution. 

Whether BHs in this mass range ($m_1\lesssim 15M_\odot$) can indeed systematically form with such large natal kicks $(\gtrsim100\,\rm km/s)$ is highly uncertain. Analytic estimates have suggested that a fallback supernova with asymmetric ejecta can, in principle, impart substantial kicks to newly formed BHs \citep{2013MNRAS.434.1355J}, and recent three-dimensional core-collapse simulations have demonstrated the possibility of this scenario \citep{2020MNRAS.495.3751C,2025ApJ...987..164B}. However, producing both a large kick and a $\sim10\,M_\odot$ BH  requires rather specific conditions: the explosion must be sufficiently energetic and asymmetric to eject enough momentum, while still allowing enough fallback to form a massive BH. Existing three-dimensional model suites contain both low-kick BH formation cases and high-kick examples with results not in agreement even for the same progenitor star \citep{2024Ap&SS.369...80J,2025ApJ...987..164B}. Therefore the frequency of such outcomes in nature remains unclear. Alternatively, a BH may form with substantial natal kicks by asymmetric loss of the convective hydrogen envelope by the mass-decrement effect \citep{Nadezhin1980,Lovegrove2013,Antoni2022}. While it remains difficult to robustly predict the BH natal kick distribution from core-collapse modeling, there is observational evidence that at least some BHs in this mass range are born with small natal kicks \citep[e.g.,][]{2024PhRvL.132s1403V,whitaker2026longperiodstellarmassblack} and a controversy about some BHs in low-mass X-ray binaries that could indicate large natal kicks $\gtrsim100\,\rm km/s$ \citep[][and references therein]{Mandel2016}.

Alternatively, the observed BBH mergers could be explained by
 the large abundance of hierarchical stellar triples or higher-order configurations \citep{Silsbee2017,Antonini2017,Grishin2018,Liu2018,Antonini2018,Rodriguez2018,Mangipudi2022,Stegmann2022,Stegmann2022b,2025A&A...699A.272V,Dorozsmai2025,Stegmann2025,2026ApJ..1000L..59S}. In these systems,  the merger of a BBH formed in the inner binary is driven by the gravitational perturbation from a distant tertiary companion that drives large-amplitude eccentricity oscillations through the Lidov-Kozai effect \citep{Zeipel1910,Lidov1962,Kozai1962}. Mergers formed through the triple channel have been shown to reproduce the inferred event rate from the LIGO-Virgo-KAGRA observations \citep[e.g.,][]{Rodriguez2018,Stegmann2025}.
 %Alternatively, the $\chi_{\rm eff}$ distribution may be recovered if the BBH mergers are driven by the Lidov-Kozai effect in hierarchical triples,

The  $\chi_{\rm eff}$ distribution expected under the triple scenario is shown in Fig.~\ref{fig:astro_model_comparison_triples}. Here, the BBH mergers are obtained from a simulation of a triple star population in which the inner binary star evolution is modeled with the same physical assumptions as in the orange model in Fig.~\ref{fig:astro_model_comparison_binaries_Aleksandra} \citep[][]{Olejak2024}. In particular, it assumes that BH natal kicks are suppressed by the standard fallback supernova mechanism. The three-body dynamics of resulting BBHs is evolved with the secular integration scheme presented in \cite{Antonini2018} and \cite{Rodriguez2018}. This model broadly recovers a peak at a small positive $\chi_{\rm eff}$ value and the imbalance between a significant contribution of $\chi_{\rm eff}<0$ and dominating contribution of $\chi_{\rm eff}>0$. The characteristic shape results from the combined action of the Lidov-Kozai effect, relativistic de Sitter precession of the BH spins, and gravitational-wave emission \citep{Antonini2018,Liu2018,Rodriguez2018,Su2021}, which enables a wide range of spin-orbit angles $-1\lesssim\cos\theta_{1(2)}\lesssim1$ at merger and a characteristic peak near $\cos\theta_{1(2)}\gtrsim0$ and hence $\chi_{\rm eff}\gtrsim0$. 

Previous work already identified the qualitative shape of the $\chi_{\rm eff}$ from triples under simplified assumptions for the BH spin magnitude \citep{Antonini2018,Liu2018,Rodriguez2018,Stegmann_inprep} and found some tentative agreement of the triple distribution with the less accurately measured distribution of $\cos\theta_{1(2)}$ \citep{2026ApJ..1000L..59S,Wolfe2026}. Here, we assume  that the BH spin magnitudes are independently drawn from a truncated normal distribution between $\chi_{\rm min}=0$ and $\chi_{\rm max}=1$ with $\mu=0.05$ and varying $\sigma$ between 0.05 and 0.20 which yield broad agreement with the $\chi_{\rm eff}$ distribution for $\sigma\lesssim0.1$. %However, a detailed investigation under which physical assumptions in the triple star evolution the resulting $\chi_{\rm eff}$ distribution fits the inferred one the most, e.g., the precise location of the peak and the fraction of negative $\chi_{\rm eff}$, is left for future work \citep{Stegmann_inprep}.

We emphasize three standard assumptions in the isolated binary and triple modeling above: (i) the progenitor stars are born with spins aligned with the (inner) orbital angular momentum vector or stellar tides are efficient enough to align them, (ii) BHs inherit the spin direction of their progenitor stars, and (iii) the spatial direction of any natal kick is random. Revising either one of the assumption would have profound implications for our understanding of massive star formation, tidal interactions between massive binary stars, and BH formation mechanisms, but may offer  alternative pathways to obtain substantial fractions of negative $\chi_{\rm eff}$ from the evolution of isolated binaries \citep{Tauris2022,BaibhavKalogera2024}.

Finally, we note that we have considered only the most commonly adopted explanation for the low-mass population, namely formation of the BH binaries in the dominant birth environment of their stellar progenitors, i.e., field binaries or triples. Other channels, including dynamical formation in dense stellar clusters \cite[e.g.,][]{Askar2017,2019PhRvD.100d3027R,2021MNRAS.505..339M}, AGN disks \cite{Bartos2016}, primordial black holes \cite[e.g.,][]{2016PhRvL.117f1101S}, or a combination of these, could also contribute to the inferred population. In particular, dynamical formation naturally produces binaries with misaligned spins and can therefore account for the significant support at negative $\chi_{\rm eff}$ inferred from the data. It is less straightforward, however, for these channels alone to reproduce the pronounced mode at positive $\chi_{\rm eff}$. A mixed population, with contributions from both field binaries/triples and dynamically assembled systems, could potentially explain both features: the field component producing the preference for positive $\chi_{\rm eff}$, while the dynamical component broadens the distribution and populates the negative-$\chi_{\rm eff}$ region. Distinguishing such a mixture from a single formation channel will require modeling these alternatives explicitly, which we leave for future work.

\section{Conclusion}
In this work, we employ single-mass transition in $\chi_{\rm eff}$ models, a flexible non-parametric model $\mathcal{GP}+\mathcal{GP}$ and a parametric model $\mathcal{N}+\mathcal{NU}$, to investigate the existence of a leading-order transition of $\chi_{\rm eff}$ in the low-mass regime. As a result, we infer the transitions $\tilde{m} = 15.1^{+4.4}_{-3.7}\,M_\odot$ and $ 15.1^{+3.8}_{-3.9}\,M_\odot$ for the $\mathcal{GP}+\mathcal{GP}$ and the $\mathcal{N}+\mathcal{NU}$ models respectively. We test the significance of our result by completing a leave-one-out analysis with GW241110 and GW241011, demonstrating that the transition is unlikely to be driven by a small subset of events. In comparing the median and width of the $\chi_{\rm eff}$ distribution above and below this transition, we find them to be  statistically distinct from one-another. 
We validate our results by performing a mock data analysis, in which we draw realizations from a $\chi_{\rm eff}$ population consistent with that inferred in GWTC-5.0, with no correlation between $\chi_{\rm eff}$ and $m_{1}$. In this mock analysis, we find that all but two realizations (100 in total) have the mass transition railing against the prior. As such, we find that our transition is unlikely to be the result of a particular realization randomly drawn from a $\chi_{\rm eff}$ distribution similar to that in GWTC-5.0.

We follow this up by looking at potential origins of the inferred transition by exploring  formation scenarios of the low mass population. We begin by demonstrating that isolated binary evolution with standard fallback kick models cannot  easily reproduce the negative $\chi_{\rm eff}$ fraction. For this scenario to reproduce the support at negative $\chi_{\rm eff}$, the black holes below $\tilde{m}$ need to be quite large ($\gtrsim100\,\rm km/s$). However, the likelihood of these BHs receiving such natal kicks is uncertain. As such, we look alternatively at hierarchical triples; here we find that this model broadly reconstructs the $\chi_{\rm eff}$ distribution inferred from the GW data analysis below $\tilde{m}$, including  the negative $\chi_{\rm eff}$ fraction.

\bigskip
\noindent {\it Acknowledgments}. J.S. and A.O. thank Hans-Thomas Janka and Daniel Kresse for very insightful discussions about the natal kick velocities at BH formation. 
FA  is supported by the UK’s Science
and Technology Facilities Council grants ST/V005618/1
and UKRI2489. IMRS acknowledges support from the
 Technology Facilities Council Ernest Rutherford Fellowship grant number UKRI2423.
This material is based upon work supported by
NSF’s LIGO Laboratory which is a major facility fully funded by
the National Science Foundation, as well as the Science and Technology Facilities Council (STFC) of the United Kingdom, the Max-Planck-Society (MPS), and the State of Niedersachsen/Germany for
support of the construction of Advanced LIGO and construction and
operation of the GEO600 detector. Additional support for Advanced
LIGO was provided by the Australian Research Council. Virgo is
funded, through the European Gravitational Observatory (EGO), by
the French Centre National de Recherche Scientifique (CNRS), the
Italian Istituto Nazionale di Fisica Nucleare (INFN) and the Dutch
Nikhef, with contributions by institutions from Belgium, Germany,
Greece, Hungary, Ireland, Japan, Monaco, Poland, Portugal, Spain.
KAGRA is supported by Ministry of Education, Culture, Sports, Science and Technology (MEXT), Japan Society for the Promotion of
Science (JSPS) in Japan; National Research Foundation (NRF) and
Ministry of Science and ICT (MSIT) in Korea; Academia Sinica
(AS) and National Science and Technology Council (NSTC) in Taiwan. This research has made use of data or software obtained from the Gravitational Wave Open Science Center (gwosc.org), a service of the LIGO Scientific Collaboration, the Virgo Collaboration, and KAGRA. The authors are grateful for computational resources provided by the LIGO-Virgo Collaboration and by Cardiff University and supported by STFC grant ST/V005618/1.

\bigskip
\noindent {\it Author contribution statement}. F.A. and J.S. conceived the study. E.F. performed the gravitational-wave population analysis.   J.S. wrote Section~4 and performed the triple simulations. I.R.-S. performed the mock-data analysis and wrote the corresponding section and appendix material. F.A. and E.F. wrote most of the remaining sections.
T.C. provided the original code on which the analysis code used in this manuscript is based, and assisted with the data analysis. A.O. assisted with the astrophysical interpretation of the results and provided the simulation data used in Fig.~3. All authors contributed to the manuscript and to the interpretation of the results.

%% file: appendix.tex
% \clearpage
% \onecolumngrid

% % Reset counters
% \setcounter{equation}{0}
% \setcounter{figure}{0}
% \setcounter{table}{0}
% \setcounter{section}{0}

% % Supplementary numbering
% \renewcommand{\theequation}{S\arabic{equation}}
% \renewcommand{\thefigure}{S\arabic{figure}}
% \renewcommand{\thetable}{S\arabic{table}}

% % Remove section numbering
% \renewcommand{\thesection}{}
% \renewcommand{\thesubsection}{}

% Title
% \begin{center}
% {\Large \bf Appendix}
% \end{center}

% \vspace{1cm}

% \begin{figure}
% \centering
% \resizebox{0.5\textwidth}{!}{\includegraphics{figs/Xeff_mainfig_2heaviGP_mCutfree_pXeff.pdf}}
% \caption{The left panel shows the transition mass scale $\tilde{m}$ posterior inferred from the non-parametric $\mathcal{N} + \mathcal{NU}$ model. The right panel contains the $p(\chi_{\rm eff})$ distribution above and below the inferred mass transition. The shaded area in the right panel show the $90 \%$ confidence interval and the solid lines are the recovered median of the distribution.}
% \label{fig:posterior}
% \end{figure}

\subsection{Hierarchical inference and data}
We analyze BBH candidates in GWTC-5 \citep{2026arXiv260527225T}. For events in GWTC-1 \citep{2019PhRvX...9c1040A}, we use the \textit{Overall posterior} samples. For GWTC-2 events \citep{Abbott_2021}, we use the \textit{PrecessingSpinIMRHM} samples, while for GWTC-3 events \citep{PhysRevX.13.041039} we use the \textit{C01:Mixed} samples. For GWTC-4 \citep{LVKcat_inprep} and GWTC-5 \citep{2026arXiv260527225T}, we use \texttt{NRSur7dq4} samples when available; otherwise, we use \textit{Mixed} samples for GWTC-4 and \textit{IMRPhenomXPHM-SpinTaylor} \cite{PhysRevD.103.104056, PhysRevD.111.104019} samples for GWTC-5.

% , restricting to events with false-alarm rate ${\rm FAR}<1,{\rm yr}^{-1}$, as in previous population studies \citep{2025arXiv250818083T}. 
% This selection yields 259 events. Selection effects are computed using recovered injections from the public injection campaigns \citep{PhysRevX.13.041039,44x3-hv3y}. 

\begin{figure*}
\centering
\resizebox{1.\textwidth}{!}{\includegraphics{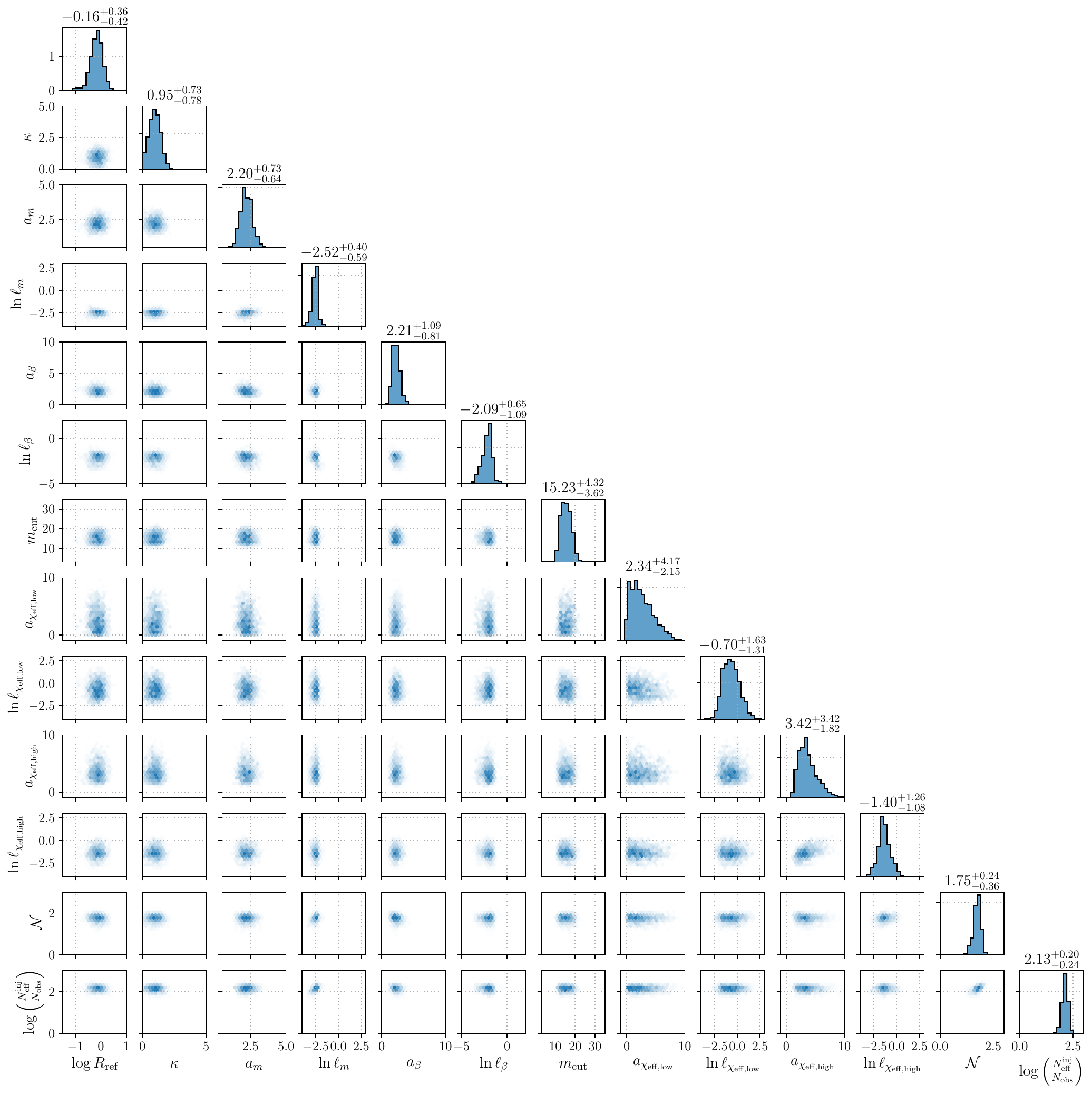}}
\caption{Posterior distributions for hyperparameters described in Table~\ref{tab:priors}. Here, log$R_{\rm ref}$ represents the differential merger-rate density, evaluated at $m_{1} = 20$ M$_{\odot}$ and $z = 0.2$ redshift, in units Gpc$^{-2}$yr$^{-1}$M$_{\odot}^{-1}$. Also, log$N^{\rm inj}_{\rm eff} / N_{\rm obs}$ is the total number of injections divided by the number of detections.
}
\label{fig:param_posteriors}
\end{figure*}

\begin{table}
\centering
\begin{tabular*}{\columnwidth}{@{\extracolsep{\fill}} l c c}
\hline
Parameter & Prior & Defined in \\ 
\hline
$\tilde{m}$ & $\mathcal{U} [3M_\odot,\, 35M_\odot]$ & $m_{1}$ Model \\
\hline
$a_{m_{1}}$ & $\mathcal{HN} (0.8)$ & $m_{1}$ Model \\
ln$\,l_{m_{1}}$ & $\mathcal{N} (-0.2, 1)$ & $m_{1}$ Model \\
\hline
$a_{\beta_{q}}$ & $\mathcal{HN} (1.0)$ & $\beta_{q}$ Model \\
ln$\,l_{\beta_{q}}$ & $\mathcal{N} (-0.5, 1))$ & $\beta_{q}$ Model \\
\hline
$a_{\chi_{\rm eff}, \, i}$ & $\mathcal{HN} (3)$ & $\chi_{\rm eff}$ Model \\
ln$\,l_{\chi_{\rm eff}, \, i}$ & $\mathcal{N} (-0.5, 0.9)$ & $\chi_{\rm eff}$ Model \\
\hline
$\chi_{max, \, i}$ & $\mathcal{U} [0.1,\, 1]$ & $\mathcal{NU}$ $\chi_{\rm eff}$ Model \\
$\chi_{max, \, i}$ & $\mathcal{U} [0.05,\, 1]$ & $\mathcal{GP}$ $\chi_{\rm eff}$ Model \\
$\chi_{min,unscaled, \, i}$ & $\mathcal{U} [0,\, 1]$ & $\chi_{\rm eff}$ Model 
\\
\hline
$\mu_i$ & $\mathcal{U}(-1)$ & $\mathcal{N}+\mathcal{U}$ 
 $\chi_{\rm eff}$ Model  \\
$\sigma_i$ & $\mathcal{U}(-1.5, \, 0)$ & $\mathcal{N}+\mathcal{U}$ 
 $\chi_{\rm eff}$ Model  \\
$\theta$ & $\mathcal{U}(0, \, 1)$ & $\mathcal{N}+\mathcal{U}$ 
 $\chi_{\rm eff}$ Model  \\
\hline
$\kappa$ & $\mathcal{N}(0, \, 6)$ & redshift Model \\
\hline
\end{tabular*}
\caption{Prior distributions for hyperparameters of $p(\chi_{\rm eff})$ and  mass models; $\mathcal{U}$, $\mathcal{N}$
and $\mathcal{HN}$ indicate uniform, normal, and half--normal distributions, respectively. As in the text, we use $i\in{{\rm low},{\rm high}}$.}
\label{tab:priors}
\end{table}

We infer the properties of the BBH population with a hierarchical Bayesian analysis implemented with Hamiltonian Monte Carlo. Sampling is performed with \texttt{numpyro} \citep{phan2019composable}, using \texttt{jax} \citep{jax2018github}, and follows the population-inference framework described in \cite{nxnr-pdyx} to which we redirect the reader for more details.

For a set of detected events with data $\{d_i\}$, the posterior on the
population hyperparameters $\Lambda$ is
\begin{equation}
p(\Lambda \mid \{d_i\})
\propto
p(\Lambda)\,
\xi^{-N_{\rm det}}(\Lambda)
\prod_{i=1}^{N_{\rm det}}
\left\langle
\frac{p(\theta_i\mid\Lambda)}
     {p_{\rm pe}(\theta_i)}
\right\rangle_i .
\label{eq:pop_likelihood}
\end{equation}
Here $p(\Lambda)$ is the hyperprior, $p_{\rm pe}(\theta_i)$ is the
single-event prior, and the brackets denote an average over posterior
samples for event $i$ \citep{nxnr-pdyx,Fishbach_2018}. Selection effects
are included through
\begin{equation}
\xi(\Lambda)
=
\frac{1}{N_{\rm inj}}
\sum_{j=1}^{N_{\rm found}}
\frac{p(\theta_j\mid\Lambda)}
     {p_{\rm inj}(\theta_j)} ,
\label{eq:selection}
\end{equation}
estimated from recovered injections.

We monitor Monte Carlo uncertainty using the effective number of samples,
\begin{equation}
N_{\rm eff}
=
\frac{\left(\sum_j w_j\right)^2}
     {\sum_j w_j^2} ,
\label{eq:neff}
\end{equation}
where $w_j$ is the corresponding reweighting factor for either event
posterior samples or injections. Following
\cite{essick2022precisionrequirementsmontecarlo}, we require
$N^{\rm inj}_{\rm eff}\gtrsim4N_{\rm det}$ and apply a smooth likelihood
penalty in regions with insufficient Monte Carlo support.

We write the differential merger rate density as
\begin{eqnarray}
R(m_1,m_2,z,\chi_{\rm eff})
=
R_{\rm ref}
\frac{f(m_1)}{f(20,M_\odot)}
\frac{(1+z)^{\kappa}}{(1.2)^\kappa}
\nonumber \\
\times
p(m_2\mid m_1)
p(\chi_{\rm eff}\mid m_1),
\label{eq:rate_density}
\end{eqnarray}
where $R_{\rm ref}$ is the merger rate density evaluated at $m_1=20,M_\odot$ and $z=0.2$. 

We model the redshift evolution as a power law in $1+z$ \citep{Fishbach_2018,Callister_2020},
\begin{equation}
p(z)
\propto
\frac{dV_c}{dz}
\frac{(1+z)^{\kappa-1}}{(1.2)^\kappa}.
\label{eq:redshift}
\end{equation}

The primary-mass distribution is modeled non-parametrically with a Gaussian process. Specifically, we write
\begin{equation}
f(m_1)
=
\exp\left[\Phi(\ln m_1)\right],
\label{eq:m1_spectrum}
\end{equation}
where
\begin{equation}
\Phi(\ln m_1)
\sim
\mathcal{GP}
\left[
0,
k(x,x';a_{m_1},l_{m_1})
\right] .
\label{eq:m1_gp}
\end{equation}

The covariance kernel is squared exponential, and the Gaussian process is evaluated on a log-spaced grid in primary mass over $2$--$200,M_\odot$.

\begin{figure*}
    \centering
    \includegraphics[width=1.0\linewidth]{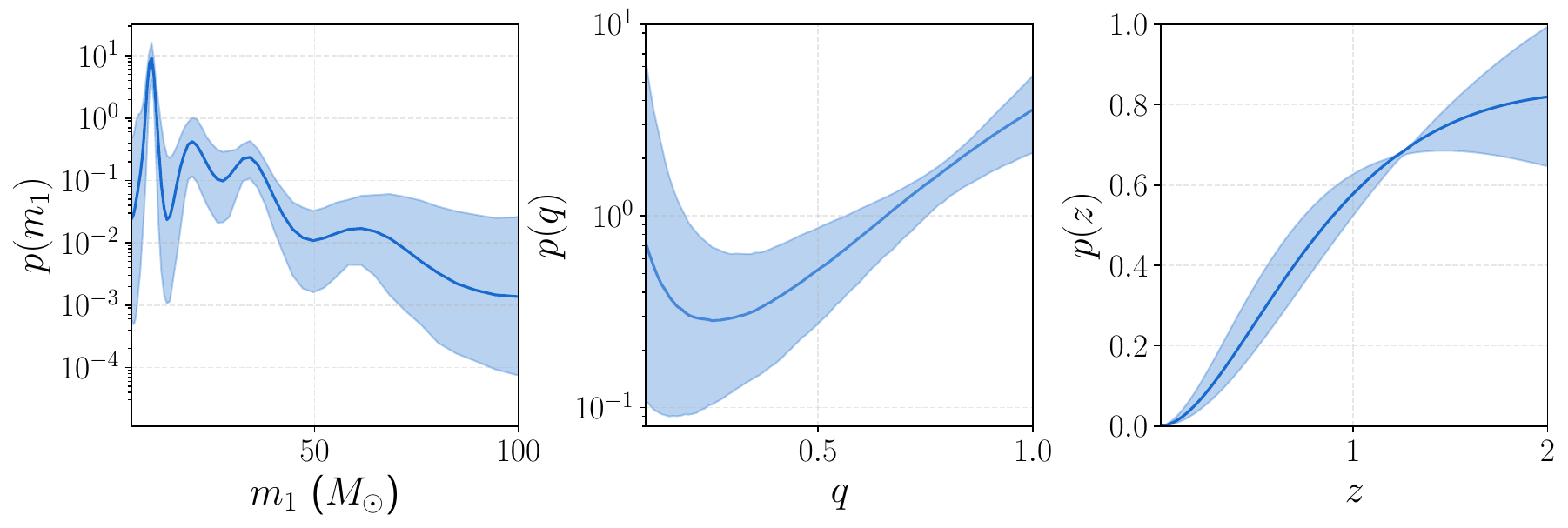}
    \caption{Inferred probability distributions of primary mass (left), mass ratio (center), and redshift (right). The spike at low $q$ is due to edge effects caused by modeling $\beta$ and $m_{1}$ non-parametrically.}
    \label{fig:pm1pqpz}
\end{figure*}

The conditional secondary-mass distribution is modeled as
\begin{equation}
p(m_2\mid m_1)
\propto
m_2^{\beta_q(m_1)} ,
\label{eq:q_model}
\end{equation}
where the mass-ratio power-law index is itself allowed to vary with primary mass,
\begin{equation}
\beta_q(m_1)=\Xi(\ln m_1),
\label{eq:betaq}
\end{equation}
with
\begin{equation}
\Xi(\ln m_1)
\sim
\mathcal{GP}
\left[
0,
k(x,x';a_{\beta_q},l_{\beta_q})
\right] .
\label{eq:betaq_gp}
\end{equation}
As for the primary-mass spectrum, the process is defined on a log-uniform grid between $2$ and $200,M_\odot$. Figure \ref{fig:pm1pqpz} shows the inferred distribution of redshift, mass ratio, and primary mass.

We model the effective-spin distribution with two non-parametric components separated by a single transition mass $\tilde m$. The conditional spin distribution is
\begin{eqnarray}
p(\chi_{\rm eff}\mid m_1)
=
p_{\rm low}(\chi_{\rm eff}),[1-\zeta(m_1)]
\nonumber \
+
p_{\rm high}(\chi_{\rm eff}),\zeta(m_1),
\label{eq:chieff_transition}
\end{eqnarray}

where $p_{\rm low}$ and $p_{\rm high}$ describe the spin distributions below and above the transition, respectively. The transition is controlled by
\begin{equation}
\zeta(m_1)
=
\frac{1}
{
1+\exp[-(m_1-\tilde m)]
},
\label{eq:zeta_single}
\end{equation}

 with $\zeta(m_1)\simeq0$ for $m_1\ll\tilde m$ and $\zeta(m_1)\simeq1$ for $m_1\gg\tilde m$. The parameter $\tilde m$ is inferred directly from the data.

Each spin component is represented by a Gaussian-process density model,
\begin{equation}
p_i(\chi_{\rm eff})
=
\frac{
\mathcal{H}_i(\chi_{\rm eff})
\exp[\Theta_i(\chi_{\rm eff})]
}{
\int_{-1}^{1}
\mathcal{H}_i(\chi_{\rm eff})
\exp[\Theta_i(\chi_{\rm eff})]
d\chi_{\rm eff}
}, 
i\in{{\rm low},{\rm high}}.
\label{eq:chieff_gp}
\end{equation}

Here $\Theta_i(\chi_{\rm eff})$ is drawn from a Gaussian process and $\mathcal{H}_i$ defines the allowed support of the distribution,

\begin{equation}
\mathcal{H}(\chi_{\rm eff})
=
\begin{cases}
1 &
\chi_{{\rm min},i}
\leq
\chi_{\rm eff}
\leq
\chi_{{\rm max},i}
\\
0 &
\text{otherwise}.
\end{cases}
\label{eq:heaviside_spin}
\end{equation}

The bounds $\chi_{{\rm min},i}$ and $\chi_{{\rm max},i}$ are inferred independently for the two spin populations. We sample the lower edge conditionally on the upper edge using

\begin{equation}
\chi_{{\rm min},i}
=
\chi_{{\rm min,unscaled},i}
\left(\chi_{{\rm max},i}+1\right)-1 .
\label{eq:chieff_bounds}
\end{equation}

This parameterization allows the data to determine both the shape and support of the effective-spin distribution on either side of the transition mass, without imposing a Gaussian, spline, or other fixed functional form. Because the density is constructed through an exponentiated Gaussian process multiplied by a variable-support window, the inferred distribution may become very small over selected regions of $\chi_{\rm eff}$ space. This flexibility is useful for identifying sharp features, extended tails, or disconnected support in the spin distribution.

The priors adopted for all hyperparameters are summarized in Table~\ref{tab:priors}. Figure~\ref{fig:param_posteriors} shows the posterior distributions for the spin model hyperparameters.

\subsection{Simulated data}

We simulate 100 realizations of 256 gravitational wave observations. These simulations are drawn from a skew-normal distribution with $\mu_\mathrm{eff}=-0.075, \sigma_\mathrm{eff}=0.17, \epsilon_\mathrm{eff}=0.9$, consistent with findings from the \textsc{Bivariate Skewed} model in the analysis of the LVK \citep{2026arXiv260527226T}. The mass model and redshift models are imported from \textsc{GWPopulation} \citep{gwpopulation} and are a \textsc{SinglePeakSmoothedMassDistribution} ($\alpha=3.14$, $m_\mathrm{min}=4.56$~M$_\odot$, $\Delta m = 5.96$~M$_\odot$, $m_\mathrm{max}=81.08$~M$_\odot$, $\beta=1.7$) and a \textsc{PowerLawRedshift} ($\kappa=2.7$). Each event posterior has 1000 samples.

We also generate $5\times10^7$ found injections from the same mass and redshift models with some different parameters ($\alpha=3$, $m_\mathrm{min}=3$~M$_\odot$, $m_\mathrm{max}=130$~M$_\odot$), and a truncated Normal distribution in $\chi_\mathrm{eff}$ ($\mu_\mathrm{eff}=0, \sigma_\mathrm{eff}=0.5, \chi_\mathrm{eff, min}=-1, \chi_\mathrm{eff, max}=1)$.

\begin{figure*}
    \centering
    \includegraphics[width=0.5\linewidth]{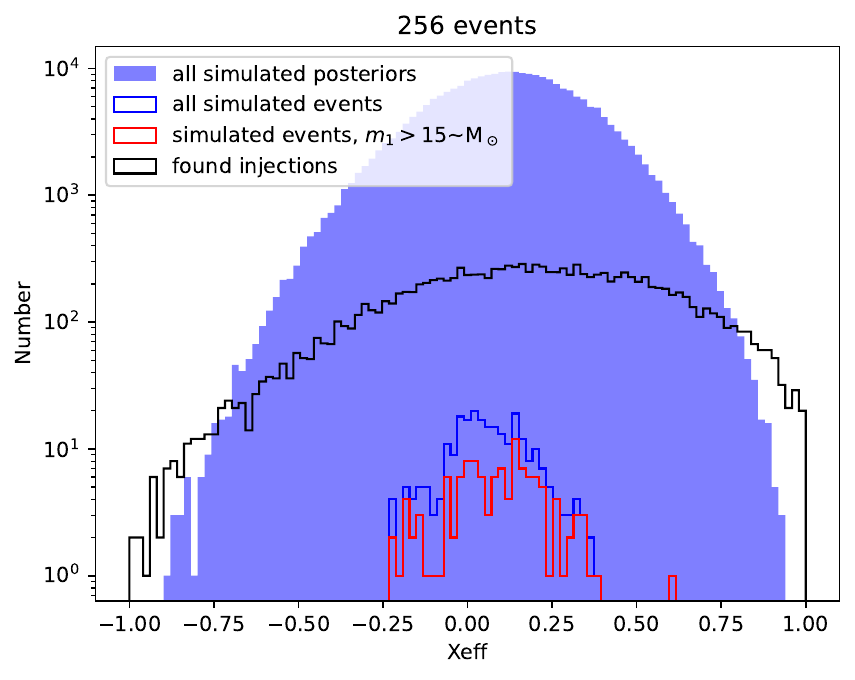}
    \caption{Simulated dataset containing mock events (blue line histogram), those events with $m_1 > 15$~M$_\odot$ (red line histogram), their simulated posterior samples (filled blue histogram), and found injections (black line histogram). The simulated dataset shown is mock population $7$, which produced the most similar results to those discussed in this paper without truly containing a transition.}
    \label{fig:simulated_data}
\end{figure*}

\begin{figure*}
    \centering
    \includegraphics[width=1.0\linewidth]{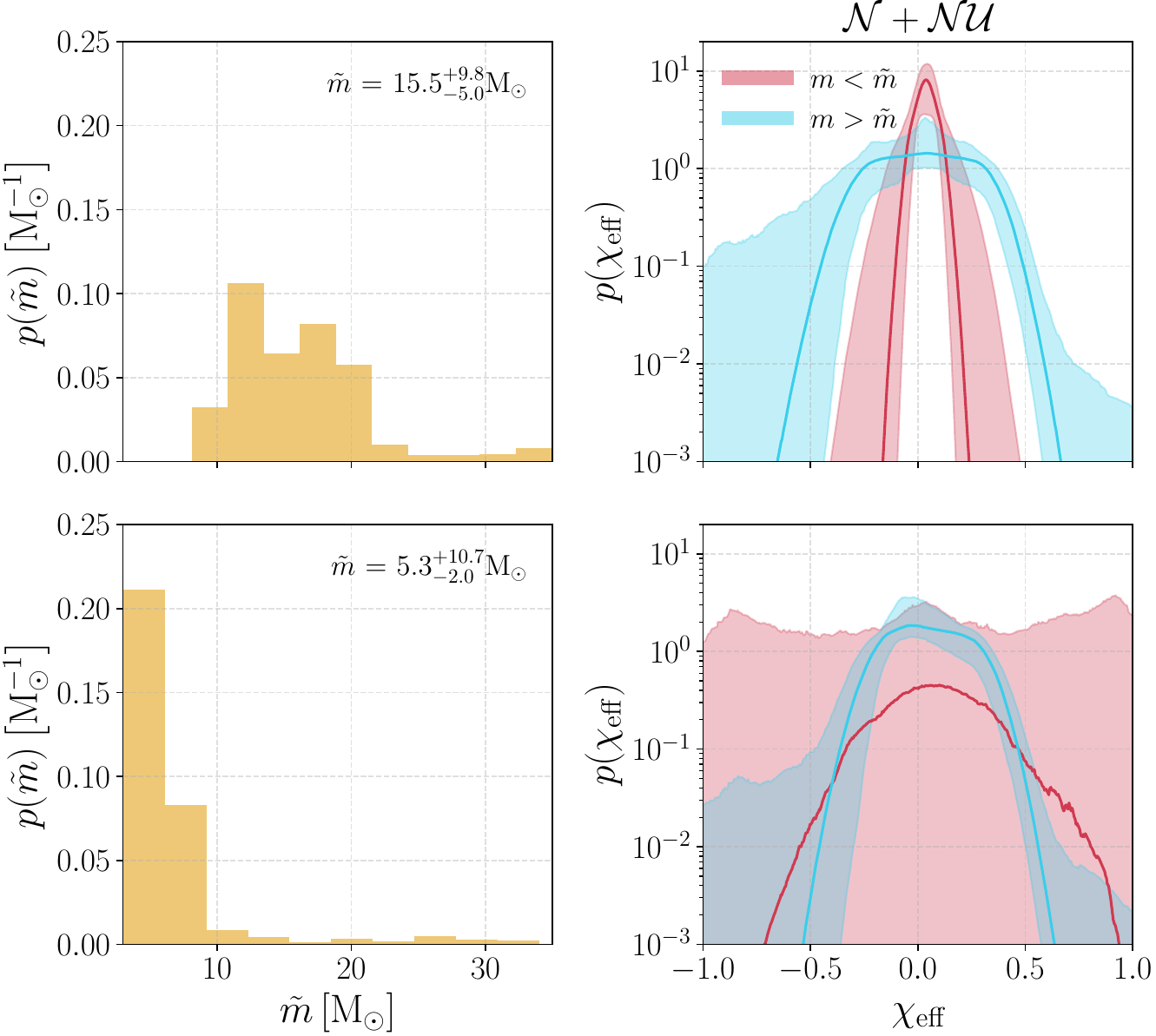}
    % \includegraphics[width=0.45\linewidth]{Xeff_1N1NU_betaGP_mCutfree_2.pdf}
    % \includegraphics[width=0.45\linewidth]{Xeff_1N1NU_betaGP_mCutfree_7.pdf}
    % ~
    % \includegraphics[width=0.45\linewidth]{Xeff_2NU_mCutfree_nomregul_mCut_posterior_2.pdf}
    % \includegraphics[width=0.45\linewidth]{Xeff_2NU_mCutfree_nomregul_mCut_posterior_7.pdf}
    \caption{The inferred mass transition $\tilde{m}$ along with the $\chi_\mathrm{eff}$ distribution when using the $\mathcal{NU}+\mathcal{U}$ model on two examples from the mock populations. Top row: a case (mock population 7) in which we recover a transition-like feature. Bottom row: the typical case where we see the posterior of $\tilde{m}$ railing against the prior.}
    \label{fig:sim_mcut_results}
\end{figure*}

In Fig.~\ref{fig:simulated_data} we show the simulated dataset used in our mock analysis, and in Fig.~\ref{fig:sim_mcut_results} we show the recovered posterior predictive distributions of
$\chi_{\rm eff}$ and of the transitional mass parameter in two cases.

\subsubsection{\textbf{Example simulated observations}}

\begin{figure}
    \centering
    \includegraphics[width=0.75\linewidth]{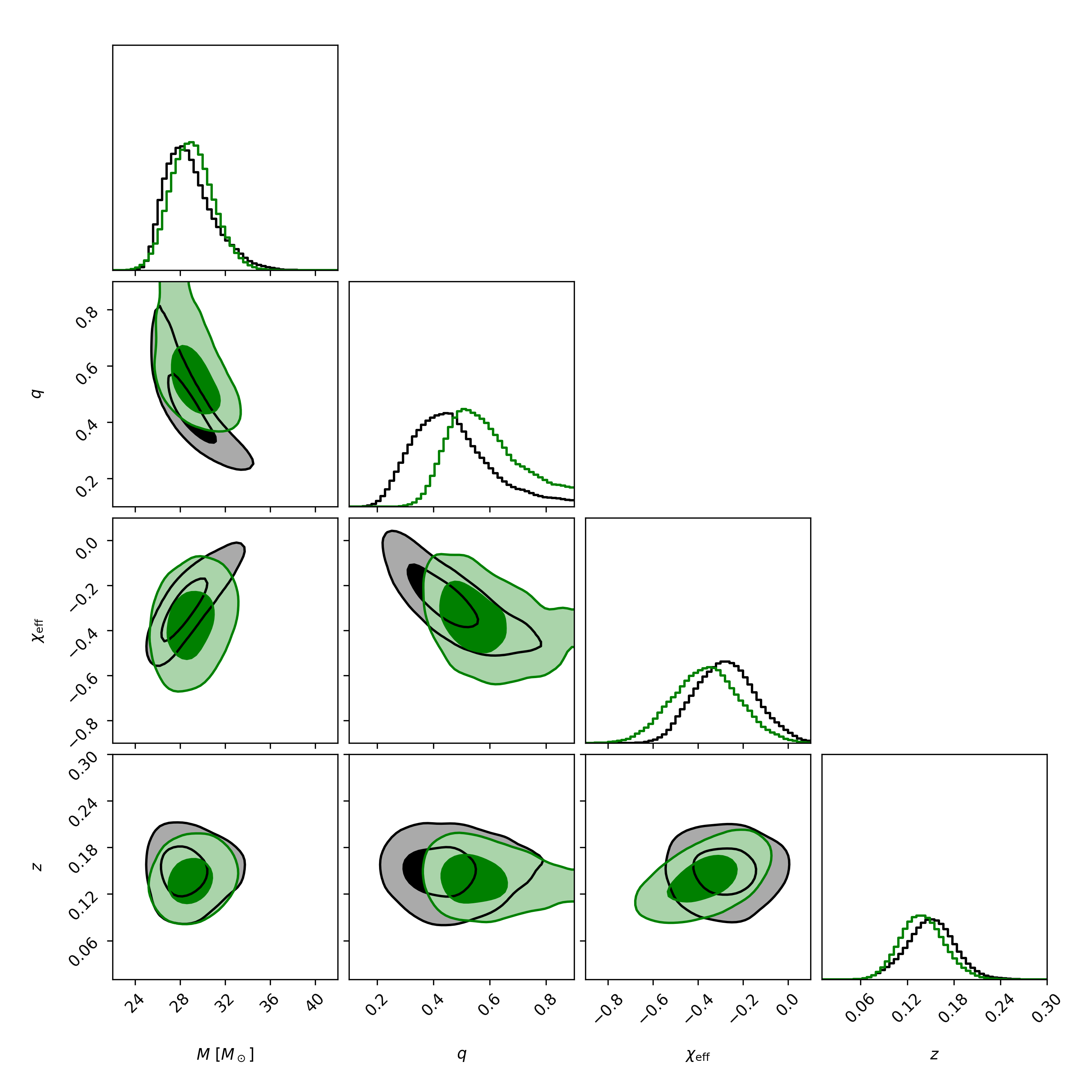}
    \caption{Example mock posterior generated with the median posterior values for GW241110 (green) compared with real GW241110 data (black).}
    \label{fig:GW241011}
\end{figure}

\begin{figure}
    \centering
    \includegraphics[width=0.75\linewidth]{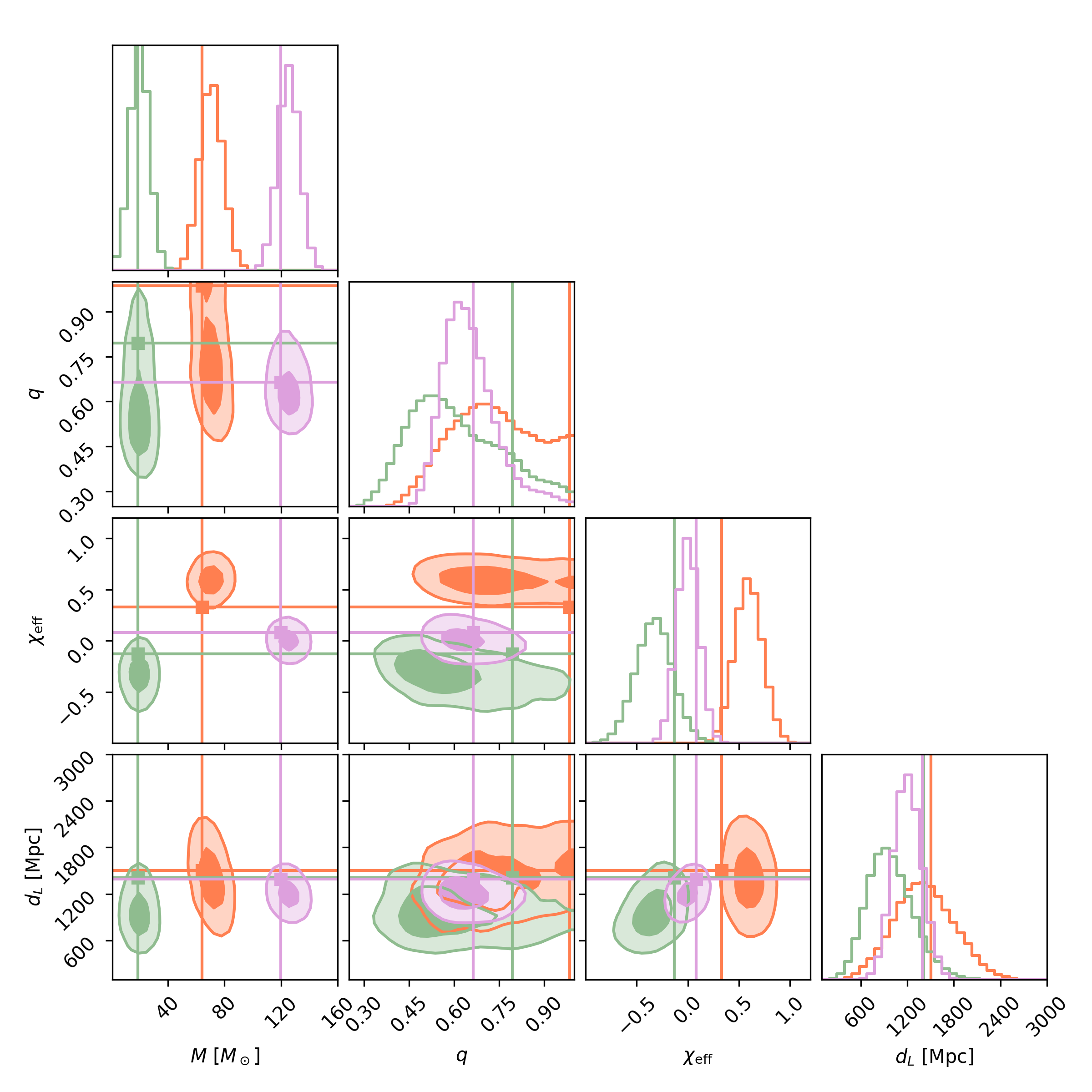}
    \caption{Three example mock observations with the most and least extreme recovered $\chi_\mathrm{eff}$ from population 7. The ``true'' injected values are indicated with vertical lines. Note that the inferred values can deviate from the injected values significantly, such that the event with the largest positive inferred $\chi_\mathrm{eff}$ is not the event with the largest positive injected $\chi_\mathrm{eff}$ ($=0.61$, a clear outlier in Fig. \ref{fig:simulated_data}).}
    \label{fig:examples}
\end{figure}

We show some examples of our simulated injections. We compare a mock posterior obtained with our default uncertainty settings with a real event, specifically, GW241110, one of the events from O4 posited as a hierarchical merger due to its unusually unequal masses and large $\chi_\mathrm{eff}$ magnitude \citep{2025ApJ...993L..21A}. We inject the median recovered $M$, $q$, $\chi_\mathrm{eff}$ and $z$ of GW241110. While the Gaussians from which uncertainties are drawn can be tweaked to more faithfully recreate this observation, we demonstrate in Figure \ref{fig:GW241011} that a draw from our default uncertainty distribution is similar to that of the real event. Notably, the correlation between $\chi_\mathrm{eff}$ and $q$ is observed in both the real and the mock data; this is due to the implementation of the ``skewed posteriors'' model described in Appendix A of \citep{2018PhRvD..98h3007N}. We additionally show three example posteriors from population 7 in Figure \ref{fig:examples}. The ability of \texttt{GWMockCat} to faithfully reproduce posteriors of real events will be explored further in \citet{IRS_inprep}.

